\documentclass[]{spie}  

\usepackage[]{graphicx}

\usepackage{psfrag}
\usepackage{amssymb}

\title{Modeling the Epps effect of cross correlations in asset prices}

\author{Bence T\'oth\supit{a,b}, B\'alint T\'oth\supit{c}, and J\'anos Kert\'esz\supit{b,d}
\skiplinehalf
\supit{a}ISI Foundation - Viale S. Severo, 65 - I-10133 Torino, Italy\\
\supit{b}Institute of Physics, Budapest University of Technology and 
Economics - Budafoki \'ut. 8. H-1111 Budapest, Hungary\\
\supit{c}Institute of Mathematics, Budapest University of Technology and 
Economics - Egry J\'ozsef u. 1. H-1111 Budapest, Hungary\\
\supit{d}Laboratory of Computational Engineering, Helsinki University of Technology 
-  P.O.Box 9203, FI-02015, Finland\\
}

\authorinfo{Bence T\'oth: E-mail: bence@maxwell.phy.bme.hu}

\begin{document} 

  \maketitle


\begin{abstract}
  We review the decomposition method of stock return cross-correlations,
  presented previously \cite{toth2007_2} for studying the dependence of the
  correlation coefficient on the resolution of data (Epps effect). Through a
  toy model of random walk/Brownian motion and memoryless renewal process (i.e. Poisson point process) of observation times we show that in case of
  analytical treatability, by decomposing the correlations we get the exact
  result for the frequency dependence. We also demonstrate that our approach produces reasonable fitting of the
  dependence of correlations on the data resolution in case of empirical data.
  Our results indicate that the Epps phenomenon is a product of the finite
  time decay of lagged correlations of high resolution data, which does not
  scale with activity. The characteristic time is due to a human time scale,
  the time needed to react to news.

\end{abstract}


\keywords{Financial correlations, Epps effect, High frequency data, Market
  microstructure, Renewal process}


\section{INTRODUCTION}
\label{sect:intro}

Stock return correlations decrease as the sampling
frequency of data increases, as reported for the first time by Epps in
1979 \cite{epps1979}. Since his discovery the phenomenon has been detected in
several studies of different stock markets
\cite{bonanno2001,zebedee2001,tumminello2006} and foreign exchange markets
\cite{lundin1999,muthuswamy2001}.

The estimation of the asymptotic cross correlations between the individual assets is of major importance since these are the main factors in classical
portfolio management. This is, however, hampered by the limited number of data. As high resulotion data are available in abundance, it is important to understand and give an accurate
description of correlations for different sampling frequencies. This is especially so,
as today the time scale in adjusting portfolios to relevant news may be
in the order of minutes. 
Since its discovery, considerable effort has been devoted to uncover the
phenomenon found by Epps
\cite{reno2003,iori2004,iori2006,kwapien2004,zhang2006,bouchaud2005}. Up to
now two main factors causing the effect have been revealed: The first one is a
possible lead-lag effect between stock returns
\cite{lo1990_1,kullmann2002,toth2006} which appears mainly between stocks
of very different capitalisation and if there is some functional dependence
between them. In this case the maximum of the time-dependent correlation
function is at non zero time lag, resulting in increasing
correlations as the sampling time scale gets into the same order of magnitude
as the characteristic lag.  This factor can be easily understood, morever, in
a recent study \cite{toth2006} we showed that through the years this effect
becomes less important as the characteristic time lag shrinks, signalising an
increasing efficiency of stock markets. It has to be emphasized that the Epps effect can also be found
in the absence of the lead-lag effect, thus in the following we will
focus only on other possible factors.

The second, more important factor is the asynchronicity of ticks in case of
different stocks \cite{reno2003,iori2004,lo1990_1,lo1990_2}. Empirical results
\cite{reno2003} showed that taking into account only the synchronous ticks
reduces to a great degree the Epps effect, i.e.  measured correlations on
short sampling time scale increase. Naturally one would expect that for a
given sampling frequency growing activity decreases the asynchronicity,
leading to a weaker Epps effect. Indeed Monte Carlo experiments showed an
inverse relation between trading activity and the correlation drop
\cite{reno2003}. 

In our previous papers \cite{toth2007,toth2007_2} we introduced a framework
for describing the correlations on different time scales. We discussed the
deficiencies of existing descriptions of the phenomenon, especially the fact
that the characteristic time of the Epps effect does not scale with activity,
thus can not be solely caused by the asynchronicity of ticks, and presented a
decomposition process of the equal-time correlations on all time scales by
writing them as functions of time dependent correlations on shorter time
scales. We demonstrated the decomposition on a model case and showed fits for
the Epps curves in case of real data, getting a good agreement with the
measured correlations. In this paper we elaborate on the toy model\cite {toth2007_2}
showing that the result through decomposing the correlations leads us to the exact solution.

In the following, first we summarize the decompostion of correlations written
in details in our previous paper (Section \ref{decomposition}). In Section
\ref{analyt} we show that the decomposition process leads to the exact
analytic solution in a treatable model case.  At the end of the paper (Section
\ref{data}) we show an example of fitting the Epps curve for real stock data
and review the process we believe to lie under the phenomenon.

\section{Decomposition of correlations}
\label{decomposition}

We are interested in correlations between the logarithmic returns of stock
prices as a function of the sampling time scale of data. The log-returns are
defined by:
\begin{eqnarray}
\label{eq:ret}
r_{\Delta t}^{A}(t)=\ln \frac{p^{A}(t)}{p^{A}(t-\Delta t)},
\end{eqnarray}
where \(p^{A}(t)\) stands for the price of stock \textit{A} at time \(t\). 
Throughout the paper we will assume that the return distributions are stationary both empirically and in the model. The time dependent correlation function \(C_{\Delta t}^{A/B}(\tau)\) of stocks
\textit{A} and \textit{B} is defined by

\begin{eqnarray}
\label{eq:C}
C_{\Delta t}^{A/B}(\tau)=\frac{\left\langle r_{\Delta t}^{A}(t)r_{\Delta
t}^{B}(t+\tau)\right\rangle - \left\langle r_{\Delta t}^{A}(t)\right\rangle
\left\langle r_{\Delta
t}^{B}(t+\tau)\right\rangle}{\sigma^{A}\sigma^{B}}.\end{eqnarray}
The notation
\(\left\langle \cdots\right\rangle\) stands for the moving time average over the
considered period:
\begin{eqnarray}\label{eq:time_ave}
\left\langle r_{\Delta t}(t)\right\rangle =\frac{1}{T-\Delta t}\sum_{i=\Delta t}^{T} r_{\Delta t}(i),
\end{eqnarray}
where time is measured in seconds and \textit{T} is the time span of the data.
The standard deviation \(\sigma\) of the returns is:
\begin{eqnarray}\label{eq:sigma}\sigma=\sqrt{\left\langle r_{\Delta
t}(t)^{2}\right\rangle - \left\langle r_{\Delta
t}(t)\right\rangle^2},
\end{eqnarray}
both for $A$ and $B$ in Equation \ref{eq:C}. The equal-time correlation coefficient is
naturally: \(\rho_{\Delta t}^{A/B}\equiv C_{\Delta t}^{A/B}(\tau=0)\).

Using the property that returns in a certain time window $\Delta t$
are mere sums of returns in smaller, non-overlapping windows $\Delta t_{0}$,
where $\Delta t$ is a multiple of $\Delta t_{0}$ and assuming the time
average of stock returns to be zero, we are able to deduce the following
relationship between correlations on different time scales (for details see
Ref. \citenum{toth2007_2}):

\begin{eqnarray}
\label{eq:data_formula2}
\rho_{\Delta t}^{A/B}=\Bigg(\sum_{x=-\frac{\Delta t}{\Delta t_0}+1}^{\frac{\Delta t}
    {\Delta t_0}-1}\left(\frac{\Delta t}{\Delta
        t_0}-|x|\right)f_{\Delta t_0}^{A/B}(x\Delta t_0)\Bigg)\times
    \nonumber \\
\Bigg(\sum_{x=-\frac{\Delta t}{\Delta t_0}+1}^{\frac{\Delta t}{\Delta t_0}-1}\left(\frac{\Delta t}{\Delta
      t_0}-|x|\right)f_{\Delta
      t_0}^{A/A}(x\Delta t_0)\Bigg)^{-1/2}\times \nonumber \\ 
\Bigg(\sum_{x=-\frac{\Delta t}{\Delta t_0}+1}^{\frac{\Delta t}
    {\Delta t_0}-1}\left(\frac{\Delta t}{\Delta
        t_0}-|x|\right)f_{\Delta t_0}^{B/B}(x\Delta t_0)\Bigg)^{-1/2}\rho_{\Delta t_0}^{A/B}.
\end{eqnarray}.

In Equation \ref{eq:data_formula2} $f_{\Delta t_0}^{A/B}(x\Delta
t_0)$, $f_{\Delta t_0}^{A/A}(x\Delta t_0)$ and $f_{\Delta t_0}^{B/B}(x\Delta
t_0)$ are the decay functions of lagged correlations on the short time scale
($\Delta t_0$) given by the expression

\begin{eqnarray}
\label{eq:def_decay}
f_{\Delta t_0}^{A/B}(x\Delta t_0)=\frac{\left\langle r_{\Delta t_0}^{A}(t)r_{\Delta t_0}^{B}(t+x\Delta t_0)\right\rangle}{\left\langle r_{\Delta t_0}^{A}(t)r_{\Delta t_0}^{B}(t)\right\rangle},
\end{eqnarray}
(and similarly for $f_{\Delta t_0}^{A/A}(x\Delta t_0)$ and $f_{\Delta
  t_0}^{B/B}(x\Delta t_0$)), defined for both positive and negative $x$ values.

This way we obtained an expression of the correlation coefficient for any
sampling time scale, $\Delta t$, by knowing the coefficient on a shorter
sampling time scale, $\Delta t_0$, and the decay of lagged correlations on the
same shorter sampling time scale (given that $\Delta t$ is multiple of $\Delta
t_0$). Our method is to measure the correlations and fit their decay functions
on a certain short time scale and compute the Epps curve using the above
formula.

\section{Analytically treatable case}
\label{analyt}

In this section we demonstrate that the solution through the decomposition of
the correlations leads to the exact solution in case of analytical tretability
of the decay functions. First we discuss a toy model describing two correlated
but asynchronous time series, then we show that the two ways of deducing
expressions for the relation of the correlations on different time scales
lead to the same result.

\subsection{The model}
\label{model}

We would like to study generated time series which have similar properties as
real world price time series.  To do this, we simulate two correlated but
asynchronous logarithmic price time series. As a first step we generate a core random walk
with unit steps up or down in each second with equal possibility ($W(t)$).
Second we sample the random walk, $W(t)$, twice independently with waiting
times drawn from an exponential distribution.  This way we obtain two time
series ($\log p^A(t)$ and $\log p^B(t)$), which are correlated since they are sampled
from the same core random walk, but the steps in the two walks are
asynchronous. The core random walk is:

\begin{eqnarray}
\label{eq:model_def1}
W(t)= W(t-1)+\varepsilon(t), \nonumber \\
\end{eqnarray}
where $\varepsilon(t)$ is $\pm 1$ with equal probability (and $W(0)$ is set
high in order to avoid negative values). We define the steps occuring in the
two asynchronous random walks respectively as $\underline{\omega}^A=\{\omega_{i}^{A}\}$
and $\underline{\omega}^B=\{\omega_{i}^{B}\}$ being two Poisson point processes on $\mathbb{R}^+$
with density $\lambda$, thus the time increments are drawn from the exponential
distribution:

\begin{eqnarray}
\label{eq:exp}
P(y)=\Bigg\{\begin{array}{ll} 
\lambda e^{-\lambda y} & \textrm{if } y \ge 0 \\
0 & y<0
\end{array}
\end{eqnarray}
with parameter $\lambda=1/60$. Between two consecutive steps the sampling walkers do
not move, thus:

\begin{eqnarray}
\label{eq:gammas}
\gamma^{A}(t):=max\{\omega_{i}^{A}: \omega_{i}^{A}<t\}\nonumber\\
\gamma^{B}(t):=max\{\omega_{i}^{B}: \omega_{i}^{B}<t\}
\end{eqnarray}
and the two walks become:

\begin{eqnarray}
\label{eq:p}
\log p^{A}(t):=W\big(\gamma^{A}(t)\big)\nonumber\\
\log p^{B}(t):=W\big(\gamma^{B}(t)\big)
\end{eqnarray}

\begin{figure}[htb!]
\begin{center}
\psfrag{time}[t][b][4][0]{time}
\psfrag{price}[b][t][4][0]{"log price"}
\psfrag{W(t)}[][][2.5][0]{W(t)}
\psfrag{pA(t)}[][][2.5][0]{$p^{A}(t)$}
\psfrag{pB(t)}[][][2.5][0]{$p^{B}(t)$}
\includegraphics[angle=-90,width=0.50\textwidth]{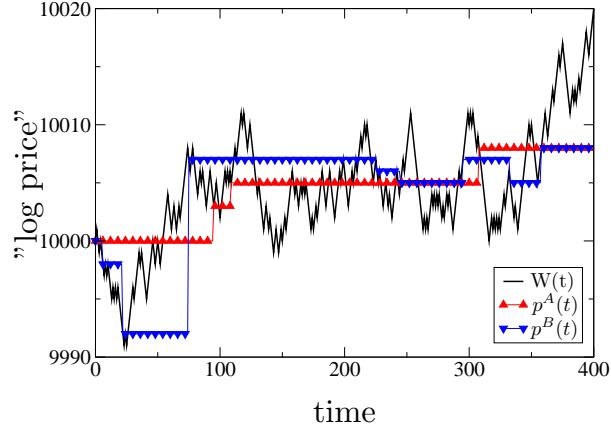}
\caption{A snapshot of the model with exponentially distributed waiting times. 
  The original random walk is shown with lines (black), the two sampled series
  (the log prices) with dots and lines (red) and triangles and lines (blue).}
\label{fig:model}
\end{center}
\end{figure}

A snapshot as an example of the generated time
series with exponentially distributed waiting times can be seen on Figure
\ref{fig:model}.

As a next step we create the return time series ($r_{\Delta t}^A(t)$ and
$r_{\Delta t}^B(t)$) of $\log p^A(t)$ and $\log p^B(t)$, and study their
cross-correlation as a function of sampling time scale. In the model case we
set the smallest time scale $\Delta t_0 =1$ time step.

\subsection{Decomposing the correlations in the model}
\label{model_deco}

\begin{figure}[htb!]
\begin{center}
\psfrag{ln(f^A/B_Dtnull(tau))}[b][t][4][0]{$\ln(f_{\Delta t_0}^{A/B}(x\Delta t_0))$}
\psfrag{tau}[t][b][4][0]{$x\Delta t_0$ [simulation steps]}
\psfrag{W(t)}[][][2.5][0]{W(t)}
\psfrag{pA(t)}[][][2.5][0]{$p^{A}(t)$}
\psfrag{pB(t)}[][][2.5][0]{$p^{B}(t)$}
\includegraphics[angle=-90,width=0.50\textwidth]{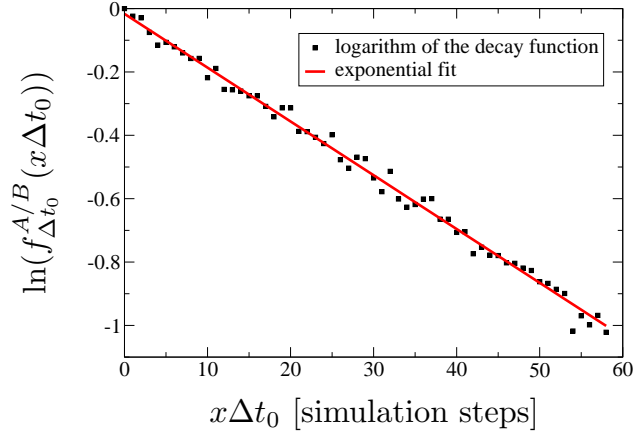}
\caption{The logarithm of the decay function and its exponential decay fit on
  a log-lin scale. The parameter of the exponential decay is $59.1$, very near
  to the parameter of the original exponential distribution of the waiting
  times.}
\label{fig:decay}
\end{center}
\end{figure}

Having a random walk model, the autocorrelation function of the
steps is zero for all non-zero time lags:

\begin{eqnarray}
\label{eq:autodecay}
f_{\Delta t_0}^{A/A}(x\Delta t_0)=f_{\Delta t_0}^{B/B}(x\Delta t_0)=\delta_{x,0}.
\end{eqnarray}
For the case when steps in the random walks are sparse in time, thus when
$\lambda\Delta t_0 \ll 1$, the decay function is an exponential
decay (see Figure \ref{fig:decay}): 

\begin{eqnarray}
\label{eq:crossdecay}
f_{\Delta t_0}^{A/B}(x\Delta
t_0)=e^{-\lambda\Delta t_0 |x|},
\end{eqnarray}
with the same parameter as the original Poisson process in Equation \ref{eq:exp}.

Thus the ratio of the correlations can be written in the following way:

\begin{eqnarray}
\label{eq:exp_anal_1}
\frac{\rho^{A/B}_{\Delta t}}{\rho^{A/B}_{\Delta t_0}}=\frac{\Delta t_0}{\Delta t}\sum_{x=-\frac{\Delta t}{\Delta t_0}+1}^{\frac{\Delta t}{\Delta t_0}-1}\left[\left(\frac{\Delta t}{\Delta
    t_0}-|x|\right)e^{-\lambda \Delta t_0 |x|}\right]=\nonumber \\
=\frac{\Delta t_0}{\Delta t}\left[ \frac{\Delta t}{\Delta
    t_0}+2\sum_{x=1}^{\frac{\Delta t}{\Delta t_0}-1}\left(\frac{\Delta
    t}{\Delta t_0}-x\right)e^{-\lambda \Delta t_0 x}\right]\nonumber \\
=1+2\sum_{x=1}^{\frac{\Delta t}{\Delta t_0}-1}e^{-\lambda \Delta t_0 x} -2\frac{\Delta t_0}{\Delta t}\sum_{x=1}^{\frac{\Delta t}{\Delta t_0}-1}xe^{-\lambda \Delta t_0 x}.
\end{eqnarray}
The first sum on the right side of Equation \ref{eq:exp_anal_1} is the sum of
a geometric series and can be written in a closed form in the following way:

\begin{eqnarray}
\label{eq:exp_anal_2}
\sum_{x=1}^{\frac{\Delta t}{\Delta t_0}-1}e^{-\lambda \Delta t_0 x}=\frac{e^{-\lambda \Delta t_0}-e^{-\lambda \Delta t}}{1-e^{-\lambda \Delta t_0}}.
\end{eqnarray}
Using the Taylor expansion of the exponential function:
\begin{eqnarray}
\label{eq:exp_anal_approx}
e^{y}=\sum_{n=0}^{\infty}\frac{y^n}{n!},
\end{eqnarray}
and applying that $\lambda \Delta t_0\ll 1$, we can neglect the high order
terms in the sum in Equation \ref{eq:exp_anal_approx} and take into account only the
terms up to linear order in $\lambda \Delta t_0$. Hence

\begin{eqnarray}
\label{eq:exp_anal_3}
\sum_{x=1}^{\frac{\Delta t}{\Delta t_0}-1}e^{-\lambda \Delta t_0 x}\approx\frac{1-\lambda
  \Delta t_0 -e^{-\lambda \Delta t}}{\lambda \Delta t_0}.
\end{eqnarray}

The second sum on the right side of Equation \ref{eq:exp_anal_1} can be obtained by differentiating \ref{eq:exp_anal_2} and taking the small $\lambda \Delta t_0$ limit:

\begin{eqnarray}
\label{eq:exp_anal_4}
\sum_{x=1}^{\frac{\Delta t}{\Delta t_0}-1}xe^{-\lambda \Delta t_0
  x}\approx 
\frac{1-\lambda \Delta t_0 +\left[ -\lambda\Delta t -1 +\lambda\Delta t_0\right]e^{-\lambda \Delta t}}{(\lambda \Delta t_0)^{2}}.
\end{eqnarray}
Inserting Equation \ref{eq:exp_anal_3} and quation \ref{eq:exp_anal_4} into Equation \ref{eq:exp_anal_1} we get:

\begin{eqnarray}
\label{eq:exp_anal_5}
\frac{\rho^{A/B}_{\Delta t}}{\rho^{A/B}_{\Delta t_0}}\approx1+\frac{2-2\lambda \Delta t_0-2e^{-\lambda \Delta
  t}}{\lambda \Delta t_0}-\nonumber\\
-\frac{2\Delta t_0}{\Delta t}\frac{1-\lambda \Delta t_0 +\left[ -\lambda\Delta t -1 +\lambda\Delta t_0\right]e^{-\lambda \Delta
  t}}{(\lambda \Delta t_0)^{2}}=\nonumber \\
=\frac{1}{(\lambda \Delta t_0)^{2}}\left[ -(\lambda \Delta t_0)^{2}+2\lambda \Delta
  t_0-\frac{2\Delta t_0}{\Delta t}+\frac{2\lambda\Delta t_0^2}{\Delta t}\right]+\nonumber \\
\frac{1}{(\lambda \Delta t_0)^{2}}e^{-\lambda \Delta t}\left( \frac{2\Delta t_0}{\Delta t}-\frac{2\lambda\Delta t_0^2}{\Delta t}\right).
\end{eqnarray}
Since $(\lambda \Delta t_0)^{2}$ and $2\lambda\Delta t_0^{2}/\Delta t$ is much
smaller than the other expressions appearing in the denominator of Equation
\ref{eq:exp_anal_5}, we can neglect them. Hence the final relation becomes

\begin{eqnarray}
\label{eq:exp_anal_6}
\frac{\rho^{A/B}_{\Delta t}}{\rho^{A/B}_{\Delta
    t_0}}\approx\frac{2}{\lambda\Delta t_0}+\frac{2}{\lambda^2 \Delta t \Delta
    t_0}\big( e^{-\lambda\Delta t}-1\big).
\end{eqnarray}

\subsection{The exact analytical solution}
\label{model_exact}

For the case described above the correlation can be given in an exact
analytical form using sepcial properties of the Poisson processes. We go to a
conrinuous description and use a Brownian motion instead of a discrete random
walk. We have:

\begin{eqnarray}
\label{eq:balint_1}
\left\langle r_{\Delta t}^{A}(t)\right\rangle=\left\langle r_{\Delta t}^{B}(t)\right\rangle=0
\end{eqnarray}
and

\begin{eqnarray}
\label{eq:balint_2}
\left\langle (r_{\Delta t}^{A}(t))^{2}\right\rangle=\left\langle (r_{\Delta
    t}^{B}(t))^{2}\right\rangle=\Delta t.
\end{eqnarray}

The interesting part of the correlation is the average of the cross-product of
the two returns, which is the following:

\begin{eqnarray}
\label{eq:balint_3}
\left\langle r_{\Delta t}^{A}(t)r_{\Delta
    t}^{B}(t))\right\rangle=\nonumber\\
=\mathbb{E}\bigg(\mathbb{E}\Big(\big(W(\gamma^{A}(t))-W(\gamma^{A}(t-\Delta t))\big)\big(W(\gamma^{B}(t))-W(\gamma^{B}(t-\Delta t))\big)\bigg|\begin{array}{ll} 
\underline{\omega}^A\\\underline{\omega}^B
\end{array}\Big)\bigg),
\end{eqnarray}
where the inner expectation averages with $\underline{\omega}^A$ and
$\underline{\omega}^B$ being given, while the outer expectation averages over
$\underline{\omega}^A$ and $\underline{\omega}^B$. Equation \ref{eq:balint_3}
can be rewritten as the expectation of the intersection of time intervals
between the last step before time $t$ and the last step before time ($t-\Delta
t$) for the two walks respectively:

\begin{eqnarray}
\label{eq:balint_4}
\left\langle r_{\Delta t}^{A}(t)r_{\Delta
    t}^{B}(t))\right\rangle=\mathbb{E}\bigg(\bigg|\Big[\gamma^{A}(t-\Delta t),\gamma^{A}(t)\Big]\cap\Big[\gamma^{B}(t-\Delta t),\gamma^{B}(t)\Big]\bigg|\bigg).
\end{eqnarray}

To detemine the expression in Equation \ref{eq:balint_4} we need to know the
probability distribution of the minimum and the maximum of two independently
and exponentially distributed variables. Let $\xi$ and $\eta$ be such. Then

\begin{eqnarray}
\label{eq:balint_5}
\mathbb{P}\big(min\{\xi,\eta\}\in(x,x+dx)\big)=2\lambda e^{-2\lambda x}dx\nonumber\\
\mathbb{P}\big(max\{\xi,\eta\}\in(x,x+dx)\big)=2\lambda (e^{-\lambda x}-e^{-2\lambda x})dx.
\end{eqnarray}
Thus the correlation coefficient becomes:

\begin{eqnarray}
\label{eq:balint_6}
\rho_{\Delta t}^{A,B}=\frac{2}{\lambda\Delta t}\int_{0}^{\lambda\Delta
  t}\big(\lambda\Delta t-x+\frac{1}{2}\big)\big(e^{-x}-e^{-2x}\big)dx=\nonumber\\
=\frac{1}{\lambda\Delta t}\big(e^{-\lambda\Delta t}-1\big)+1.
\end{eqnarray}

The ratio between the correlation coefficient on the sampling scale $\Delta t$
and sampling scale $\Delta t_0$ is

\begin{eqnarray}
\label{eq:balint_7}
\frac{\rho_{\Delta t}^{A,B}}{\rho_{\Delta t_0}^{A,B}}=\frac{\frac{1}{\lambda\Delta t}\big(e^{-\lambda\Delta t}-1\big)+1}{\frac{1}{\lambda\Delta t_0}\big(e^{-\lambda\Delta t_0}-1\big)+1},
\end{eqnarray}
which in the $\lambda\Delta t_0 \ll 1$ limit follows as

\begin{eqnarray}
\label{eq:balint_8}
\frac{\rho_{\Delta t}^{A,B}}{\rho_{\Delta t_0}^{A,B}}=\frac{2}{\lambda\Delta t_0}+\frac{2}{\lambda^2 \Delta t \Delta
    t_0}\big( e^{-\lambda\Delta t}-1\big).
\end{eqnarray}
Hence we end up with exactly the same expression as deduced through the
decomposition process in Equation \ref{eq:exp_anal_6}.

\section{Results for stock data} 
\label{data}

With the results derived in the last section we showed for a case when the
correlation can be computed analytically that our approach reproduces the
exact solution.  After this we show an example of fitting the measured
correlation on real world data with the method of decomposing the correlation
coefficient.  More examples and details can be found in Ref. \citenum{toth2007_2}.

In the analysis of real world data we used the Trade and Quote (TAQ) Database
of the New York Stock Exchange (NYSE) for the period of 4.1.1993 to
31.12.2003, containing tick-by-tick data. To avoid problems occurring from
splits in the prices of stocks, which cause large logarithmic return values in
the time series, we applied a filtering procedure. In high-frequency data, we
omitted returns larger than the $5\%$ of the current price of the stock. This
retains all logarithmic returns caused by simple changes in prices but
excludes splits which are usually half or one third of the price. We computed
correlations for each day separately and averaged over the set of days, this
way avoiding large overnight returns and trades out of the market opening
hours.

\begin{figure}[htb!]
\begin{center}
\psfrag{Dt}[t][b][4][0]{$\Delta t$ [sec]}
\psfrag{rho^MRK/JNJ_Dt}[b][t][4][0]{$\rho^{MRK/JNJ}_{\Delta t}$}
\includegraphics[angle=-90,width=0.50\textwidth]{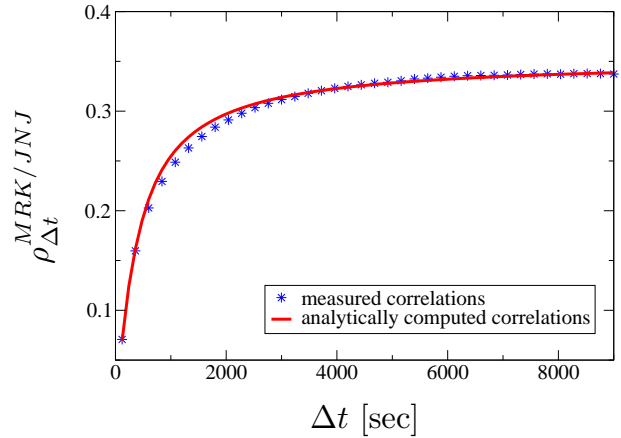}
\caption{The measured and the analytically computed correlation coefficients
  as a function of sampling time scale for the pair: MRK/JNJ. Note that using
  only the correlations measured on the smallest time scale ($\Delta t_0=120$
  seconds) we are able to give reasonable fits to the correlations on all time
  scales.}
\label{fig:fit1}
\end{center}
\end{figure}

To avoid new parameters in the model we use the raw decay functions in
Equation \ref{eq:data_formula2}, without fitting them. Since it is an
empirical approach to determine the decay functions for real data, we have to
distinguish the signal from the noise in the decay functions.  According to
this we use the decay functions for correlations only for short
time lags. For the decay of the cross-correlations we take into account the
function only up to the time lag where the decaying signal reaches zero for
the first time, for larger lags we assume it to be zero.  For the decay of
autocorrelations consider the functions only up to the time lag
where after the negative overshoot at the beginning they reach to zero from
below for the first time, for larger lags we again define them as zero.  In case of
all stock pairs studied we found the decay functions disappearing after 5--15 minutes.
In the empirical decays measured, $\Delta t_0$ is set to 2 minutes. Figure
\ref{fig:fit1} shows the measured and the analytically computed Epps curves for the
stockpair Merck \& Co., Inc. (MRK) / Johnson \& Johnson (JNJ), giving good
agreement between the measured and computed coefficients.

One can see, that the fits are able to describe the change of correlation with
increasing sampling time scale.  Through the decomposition process of the
correlations in Equation \ref{eq:data_formula2} we can see that the important
property that causes the Epps effect is the finite decay of correlations on
the high resolution scale ($\Delta t_0$). If these decays were very prompt,
the Epps phenomenon would disappear after a few seconds or minutes. This
finite decay of the correlations on the short time scale ($\Delta t_0$) is a consequence of the market
microstructure.  Reaction to a certain piece of news is usually spread out on
an interval of a few minutes for the stocks \cite{dacorogna_book,almeida1998}
due to human trading nature, thus not scaling with activity, with ticks being
distributed more or less randomly.  This means that correlated returns are
spread out for this interval (asynchronously), causing non zero lagged
correlations on the short time scale and thus the Epps effect.  This way, as
stated by Ref.  \citenum{reno2003}, the asynchronicity is indeed important in
describing the Epps effect but only in promoting the lagged correlations.
Even in case of completely synchronous, but randomly spread ticks we could
have the finite decay of lagged correlations on short time scale, and hence
the Epps effect.

\acknowledgments     
Support by OTKA T049238 and OTKA K60708 is acknowledged.





\end{document}